\documentclass[aps,prl,twocolumn,superscriptaddress,showpacs,preprintnumbers,amsmath,amssymb]{revtex4}
\usepackage{graphicx} % Include figure files
\usepackage{dcolumn}  % Align table columns on decimal point
\usepackage{amsmath}
\usepackage{epsfig}
\usepackage{xspace}
\usepackage{multirow}
\usepackage{color}
\definecolor{Bisque}{rgb}{.996,.891,.755}
\definecolor{Red}{rgb}{.996,.0,.0}
\definecolor{Blue}{rgb}{.0,.746,.996}
\definecolor{Green}{rgb}{.0,.996,.496}
\definecolor{Magenta}{rgb}{.996,.0,.996}
\renewcommand{\arraystretch}{1.1}
\def\myspecial#1{}                   %%  to print official version
\begin{document}

%%%%%%%%%%%%%%%%%%%%%%%%%%%%%%%%%%%%%%%%%%%%%%%%%%%%%%%%%%%%%%%%%%%%%%%%
%\preprint{\vbox{ \hbox{   }
%		 \hbox{Belle Preprint 2008-22}
%		 \hbox{KEK   Preprint 2008-22}
%}}
%%%%%%%%%%%%%%%%%%%%%%%%%%%%%%%%%%%%%%%%%%%%%%%%%%%%%%%%%%%%%%%%%%%%%%%%
 
\title{ \quad\\[1.0cm] \Large 
Measurement of $B^0\to\pi^+\pi^-\pi^+\pi^-$ Decays and 
Search for $B^0\to\rho^0\rho^0$
}

\tighten

\affiliation{Budker Institute of Nuclear Physics, Novosibirsk}
\affiliation{Chiba University, Chiba}
\affiliation{University of Cincinnati, Cincinnati, Ohio 45221}
\affiliation{Justus-Liebig-Universit\"at Gie\ss{}en, Gie\ss{}en}
\affiliation{The Graduate University for Advanced Studies, Hayama}
\affiliation{Gyeongsang National University, Chinju}
\affiliation{Hanyang University, Seoul}
\affiliation{University of Hawaii, Honolulu, Hawaii 96822}
\affiliation{High Energy Accelerator Research Organization (KEK), Tsukuba}
\affiliation{Institute of High Energy Physics, Chinese Academy of Sciences, Beijing}
\affiliation{Institute of High Energy Physics, Vienna}
\affiliation{Institute of High Energy Physics, Protvino}
\affiliation{Institute for Theoretical and Experimental Physics, Moscow}
\affiliation{J. Stefan Institute, Ljubljana}
\affiliation{Kanagawa University, Yokohama}
\affiliation{Korea University, Seoul}
\affiliation{Kyungpook National University, Taegu}
\affiliation{\'Ecole Polytechnique F\'ed\'erale de Lausanne (EPFL), Lausanne}
\affiliation{Faculty of Mathematics and Physics, University of Ljubljana, Ljubljana}
\affiliation{University of Maribor, Maribor}
\affiliation{University of Melbourne, School of Physics, Victoria 3010}
\affiliation{Nagoya University, Nagoya}
\affiliation{National Central University, Chung-li}
\affiliation{National United University, Miao Li}
\affiliation{Department of Physics, National Taiwan University, Taipei}
\affiliation{H. Niewodniczanski Institute of Nuclear Physics, Krakow}
\affiliation{Nippon Dental University, Niigata}
\affiliation{Niigata University, Niigata}
\affiliation{University of Nova Gorica, Nova Gorica}
\affiliation{Osaka City University, Osaka}
\affiliation{Osaka University, Osaka}
\affiliation{Panjab University, Chandigarh}
\affiliation{Saga University, Saga}
\affiliation{University of Science and Technology of China, Hefei}
\affiliation{Seoul National University, Seoul}
\affiliation{Sungkyunkwan University, Suwon}
\affiliation{University of Sydney, Sydney, New South Wales}
\affiliation{Toho University, Funabashi}
\affiliation{Tohoku Gakuin University, Tagajo}
\affiliation{Department of Physics, University of Tokyo, Tokyo}
\affiliation{Tokyo Institute of Technology, Tokyo}
\affiliation{Tokyo Metropolitan University, Tokyo}
\affiliation{Tokyo University of Agriculture and Technology, Tokyo}
\affiliation{Virginia Polytechnic Institute and State University, Blacksburg, Virginia 24061}
\affiliation{Yonsei University, Seoul}
  \author{C.-C.~Chiang}\affiliation{Department of Physics, National Taiwan University, Taipei} % Taiwan
  \author{I.~Adachi}\affiliation{High Energy Accelerator Research Organization (KEK), Tsukuba} % KEK
  \author{H.~Aihara}\affiliation{Department of Physics, University of Tokyo, Tokyo} % Tokyo
  \author{K.~Arinstein}\affiliation{Budker Institute of Nuclear Physics, Novosibirsk} % BINP
  \author{V.~Aulchenko}\affiliation{Budker Institute of Nuclear Physics, Novosibirsk} % BINP
  \author{T.~Aushev}\affiliation{\'Ecole Polytechnique F\'ed\'erale de Lausanne (EPFL), Lausanne}\affiliation{Institute for Theoretical and Experimental Physics, Moscow} % ITEP
  \author{A.~M.~Bakich}\affiliation{University of Sydney, Sydney, New South Wales} % Sydney
  \author{I.~Bedny}\affiliation{Budker Institute of Nuclear Physics, Novosibirsk} % BINP
  \author{V.~Bhardwaj}\affiliation{Panjab University, Chandigarh} % Panjab
  \author{U.~Bitenc}\affiliation{J. Stefan Institute, Ljubljana} % Ljubljana
% ...
% BINP
\collaboration{The Belle Collaboration} 

\begin{abstract}
We report on a search for the decay $B^0\to\rho^0\rho^0$ 
and other charmless modes with a $\pi^+\pi^-\pi^+\pi^-$ 
final state, 
including $B^0\to\rho^0\pi^+\pi^-$, 
non-resonant $B^0\to 4\pi^{\pm}$, 
$B^0\to\rho^0f_0(980)$, 
$B^0\to f_0(980)f_0(980)$ and 
$B^0\to f_0(980)\pi^+\pi^-$. 
These results are obtained from a data sample containing 
657 million $B \overline B$ pairs collected with the Belle 
detector at the KEKB asymmetric-energy $e^+e^-$ collider. 
We set an upper limit on $\mathcal{B}(B^0\to\rho^0\rho^0)$ of 
$1.0\times 10^{-6}$ at the 90\% confidence level (C.L.).
From our $B^0\to\rho^0\rho^0$ measurement and an isospin analysis, 
we determine the Cabibbo-Kobayashi-Maskawa phase $\phi_2$ to be 
$91.7 \pm 14.9$ degrees. 
We find excesses in $B^0\to \rho^0\pi^+\pi^-$ and 
non-resonant $B^0\to 4\pi^{\pm}$ with 1.3$\sigma$ 
and 2.5$\sigma$ significance, respectively. 
The corresponding branching fractions are less than 
$12.0 \times 10^{-6}$ and $19.3 \times 10^{-6}$ at 
the 90\% C.L.
In addition, we set 90\% C.L. upper limits as 
follows: $\mathcal{B}(B^0\to\rho^0f_0(980))< 0.3 \times 10^{-6}$, 
$\mathcal{B}(B^0\to f_0(980)f_0(980))< 0.1 \times 10^{-6}$, and 
$\mathcal{B}(B^0\to f_0(980)\pi^+\pi^-)< 3.8 \times 10^{-6}$.
\end{abstract}


\maketitle
\tighten
\normalsize

In the Standard Model (SM), $CP$ violation in the weak interaction 
can be described by an irreducible complex phase in the 
three-generation Cabibbo-Kobayashi-Maskawa (CKM) quark-mixing 
matrix \cite{1}. 
Measurements of the differences between $B$ and $\overline B$ meson 
decays provide an opportunity to determine the elements of the CKM 
matrix and thus test the SM. 
One can extract the CKM phase 
$\phi_2 \equiv \mathrm{arg} [ -(V_{td}V_{tb}^*)/(V_{ud}V_{ub}^*) ]$ 
from the time-dependent $CP$ asymmetry for the decay of a neutral 
$B$ meson via a $b\to u$ process into a $CP$ eigenstate. 
However, in addition to the $b \to u$ process, there are $b \to d$ 
penguin transitions that shift the $\phi_2$ value by $\delta \phi_2$ 
in the time-dependent $CP$ violating parameter measurement.
The shift $\delta \phi_2$ can be determined from an isospin analysis 
\cite{2} of $B\to\pi\pi$ \cite{202} or $B\to\rho\rho$ \cite{203} 
decays, or from a time-dependent Dalitz plot analysis of 
$B\to\rho\pi$ \cite{201} decays. 

For $B\to\rho\rho$ decays, 
polarization measurements in $B \to\rho^+ \rho^-$ \cite{203} and 
$B^{\pm}\to\rho^{\pm} \rho^0$ \cite{rho0rhop} show the dominance of 
longitudinal polarization, indicating that the final state in 
$B \to \rho^+ \rho^-$ is very nearly a $CP$ eigenstate. 
Measurements of the branching fraction, polarization and $CP$-violating 
parameters in $B^0\to\rho^0\rho^0$ decays complete the isospin triangle. 
The tree contribution to $B^0\to \rho^0\rho^0$ is color-suppressed, 
so its branching fraction is expected to be much smaller than that for 
$B \to\rho^+ \rho^-$ or $B^{\pm}\to\rho^{\pm} \rho^0$. 
This also makes it especially sensitive to the penguin amplitude, and 
using the $B^0\to\rho^0\rho^0$ branching fraction in an isospin analysis 
allows one to determine $\phi_2$ free of uncertainty from penguin 
contributions.

Predictions for $B^0\to\rho^0\rho^0$ using perturbative QCD 
(pQCD) \cite{32} or QCD factorization \cite{13, 14} approaches suggest 
that the branching fraction $\mathcal{B}(B^0\to\rho^0\rho^0)$ is 
at or below $1 \times 10^{-6}$, and that its longitudinal polarization 
fraction $f_{\mathrm{L}}$ is around 0.85. 
A non-zero branching fraction for $B^0\to\rho^0\rho^0$ has been 
reported by the BaBar collaboration \cite{11, 1100}; they measured 
$\mathcal{B}(B^0\to\rho^0\rho^0)=(0.92\pm 0.32\pm 0.14)\times 10^{-6}$ 
with a significance of 3.1 standard deviations ($\sigma$), 
and a longitudinal polarization fraction, 
$f_{\mathrm{L}}=0.75^{+0.11}_{-0.14} \pm 0.05$. 
They do not observe a non-resonant $B^0\to 4\pi^{\pm}$ or 
$B^0\to\rho^0\pi^+\pi^-$ contribution. 
The theoretical prediction for the non-resonant $B^0\to 4\pi^{\pm}$ 
branching fraction is around $1 \times 10^{-4}$ \cite{34}. 
The most recent measurement of this decay was made by the DELPHI 
collaboration \cite{33}, which sets a 90\% C.L. upper 
limit on the branching fraction of $2.3 \times 10^{-4}$. 

The data sample used in the analysis reported here 
contains 657 million $B \overline B$ pairs 
collected with the Belle detector at the KEKB asymmetric-energy $e^+e^-$ 
(3.5 and 8 GeV) collider \cite{101}, operating at the $\Upsilon(4S)$ 
resonance. The Belle detector \cite{102, 103} is a large-solid-angle magnetic 
spectrometer that consists of a silicon vertex detector, a 50-layer central 
drift chamber (CDC), an array of aerogel threshold Cherenkov counters (ACC), 
a barrel-like arrangement of time-of-flight scintillation counters (TOF), and 
an electromagnetic calorimeter comprised of CsI(Tl) crystals (ECL) located 
inside a superconducting solenoid coil that provides a 1.5 T magnetic field. 
An iron flux-return located outside of the coil is instrumented to detect 
$K^{0}_{L}$ mesons and to identify muons. 
Signal Monte Carlo (MC) is generated with EVTGEN \cite{Lange}, 
in which final-state radiation is taken into account with the PHOTOS 
package \cite{105}, and processed through a full detector simulation program 
based on GEANT3 \cite{Brun}.

$B^0$ meson candidates are reconstructed from neutral 
combinations of four charged pions. 
Charged track candidates are required to have a 
distance-of-closest-approach to the interaction 
point (IP) of less than 2 cm in the direction along the  
the positron beam ($z$-axis) and less than 0.1 cm in 
the transverse plane; they are also required to have a 
transverse momentum $p_T>0.1$ GeV/$c$ in the laboratory frame.  
Charged pions are identified using particle identification (PID) 
information obtained from the CDC ($dE/dx$), the ACC and the TOF. 
We distinguish charged kaons and pions using a likelihood ratio 
$\mathcal{R}_{\mathrm{PID}}=\mathcal{L}_{K}/(\mathcal{L}_{K}+
\mathcal{L}_{\pi})$, where $\mathcal{L}_{\pi}(\mathcal{L}_{K})$ 
is the likelihood value for the pion (kaon) hypothesis. We require 
$\mathcal{R}_{\mathrm{PID}} < 0.4$ for the four charged pions. 
We require that charge tracks have a laboratory momentum in 
the range [0.5, 4.0] GeV/$c$, and a polar angle in the range 
$[32.2, 127.2]^{\circ}$. 
For such tracks the pion identification efficiency is 90\%, 
and the kaon misidentification probability is 12\%. 
Charged particles that are positively identified 
as an electron or a muon are removed.

To veto $B\to D^{(*)}\pi$ and $B\to D_s \pi$ backgrounds, we remove 
candidates that satisfy either of the conditions 
$|M(h^{\pm}\pi^{\mp}\pi^{\mp})-m_{D_{(s)}}|<13 \ \mathrm{MeV}/c^2$ 
or $|M(h^{\pm}\pi^{\mp})-m_{D^0}|<13 \ \mathrm{MeV}/c^2$, 
where $h^{\pm}$ is either a pion or a kaon, and $m_{D_{(s)}}$ and 
$m_{D^0}$ are the masses of the $D_{(s)}$ and $D^0$ mesons, respectively. 
Furthermore, to reduce the $B^0\to a_1^\pm \pi^\mp$ feeddown in the 
signal region, we require that the pion with the highest momentum 
have a momentum in the $\Upsilon(4S)$ center-of-mass (CM) frame 
within the range [1.30, 2.65] GeV/$c$.

The signal event candidates are characterized by two 
kinematic variables: the beam-energy-constrained mass, 
$M_{\mathrm{bc}}=\sqrt{E^2_{\mathrm{beam}}-P^2_{B}}$, 
and the energy difference, 
$\Delta E = E_{B}-E_{\mathrm{beam}}$, where $E_{\mathrm{beam}}$ is the 
run-dependent beam energy, and $P_B$ and $E_B$ 
are the momentum and energy 
of the $B$ candidate in the $\Upsilon(4S)$ CM frame. 
In $B^0 \to \rho^0\rho^0 \to (\pi^+\pi^-)(\pi^+\pi^-)$ decays, 
or other charmless modes with a $\pi^+\pi^-\pi^+\pi^-$ final state, 
the invariant masses $M(\pi^+\pi^-)$ \emph{vs.}\ $M(\pi^+\pi^-)$ 
are used to distinguish different modes. 
There are two possible combinations for 
$M(\pi^+\pi^-)$ \emph{vs.}\ $M(\pi^+\pi^-)$: 
$(\pi^+_1\pi^-_1)(\pi^+_2\pi^-_2)$ and $(\pi^+_1\pi^-_2)(\pi^+_2\pi^-_1)$, 
where the subscripts label the momentum ordering, 
\emph{i.e.}\ $\pi^+_1$($\pi^-_1$) has a 
higher momentum than $\pi^+_2$($\pi^-_2$). 
Here we consider both $(\pi^+_1\pi^-_1)(\pi^+_2\pi^-_2)$ and 
$(\pi^+_1\pi^-_2)(\pi^+_2\pi^-_1)$ combinations and select 
candidate events if either one of the combined masses 
lies within the signal window [0.55, 1.7] GeV/$c^2$. 
This signal window is chosen to accept $\rho^0 \to \pi^+\pi^-$, 
$f_0(980) \to \pi^+\pi^-$, and non-resonant modes, and to 
exclude $K_s^0 \to \pi^+\pi^-$ and charm meson decays such 
as $D^0 \to \pi^+\pi^-$. 
If both $(\pi^+_1\pi^-_1)(\pi^+_2\pi^-_2)$ and 
$(\pi^+_1\pi^-_2)(\pi^+_2\pi^-_1)$ combinations of a candidate 
has a $\pi^+\pi^-$ pair with an invariant mass in the signal window, 
we select the $(\pi^+_1\pi^-_2)(\pi^+_2\pi^-_1)$ combination. 
According to MC simulation, this criteria selects the correct 
combination for $\rho^0\rho^0$ signal decays 98\% of the time. 
For fitting, we symmetrize the 
$M^2(\pi^+\pi^-)$ \emph{vs.}\ $M^2(\pi^+\pi^-)$ Dalitz plot 
by plotting the $\pi^+_2\pi^-_1$ $(\pi^+_1\pi^-_2)$ combination 
against the horizontal axis for events with an even (odd) 
event identification number, which is the location of the 
event in the data.

The dominant background comes from continuum 
$e^+e^- \to q\bar{q}$~($q=u,d,c~{\rm or}~s$) events. 
To distinguish signal from the jet-like continuum background, 
we use modified Fox-Wolfram moments \cite{104}, which are combined 
into a Fisher discriminant. 
This discriminant is combined with PDFs for the cosine of the $B$ 
flight direction in the CM frame and the distance in the $z$-axis 
between two $B$ mesons to form a likelihood ratio 
$\mathcal{R}=\mathcal{L}_{s}/(\mathcal{L}_{s}+\mathcal{L}_{q\overline q})$. 
Here, $\mathcal{L}_{s}$ ($\mathcal{L}_{q\overline q}$) is a likelihood
function for signal (continuum) events that is obtained
from the signal MC simulation (events in the sideband region 
$M_{\rm bc}<5.26$~GeV/$c^2$). 
We also use a flavor tagging quality variable $r$ 
provided by the Belle tagging algorithm~\cite{112} that identifies
the flavor of the accompanying $B^0$ meson in the
$\Upsilon(4S)\to B^0\overline B^0$ decay.
The variable $r$ ranges from $r=0$ for no flavor discrimination 
to $r=1$ for unambiguous flavor assignment, and it is used to divide 
the data sample into six $r$ bins.
Since the discrimination between signal and continuum events
depends on the $r$-bin, we impose different requirements
on $\mathcal{R}$ for each $r$-bin.
We determine the $\mathcal{R}$ requirement such that it maximizes a 
figure-of-merit $N_s / \sqrt{N_s + N_{q\overline q}}$,
where $N_s$ $(N_{q\overline q})$ is the expected number of
signal (continuum) events in the signal region. 
For 22\% of the events, there are multiple candidates; 
for these events we select the candidate with the smallest 
$\chi^2$ value for the $B^0$ decay vertex reconstruction. 
This selects the correct combination 79.6\% of the time.  
The detection efficiency for the signal 
is calculated by MC to be 9.16\% (11.25\%) 
for longitudinal (transverse) polarization. 
Since longitudinally polarized $B^0\to\rho^0\rho^0$ decays 
produce low momentum pions from one or both $\rho^0$'s, 
their detection efficiency is lower than that for 
transversely polarized decays.

Since there are large overlaps between $B^0\to\rho^0\rho^0$ 
and other signal decay modes in the 
$M_1(\pi^+\pi^-)$ \emph{vs.}\ $M_2(\pi^+\pi^-)$ 
distribution, we distinguish these modes by fitting to 
a large $M_1(\pi^+\pi^-)$ \emph{vs.}\ $M_2(\pi^+\pi^-)$ region. 
The signal yields are extracted by performing 
extended unbinned maximum likelihood (ML) fits.
In the fits, we use four-dimensional 
($M_{\mathrm{bc}}$, $\Delta E$, $M_1$, $M_2$) information 
to discriminate among 
$\rho^0\rho^0$, $\rho^0\pi^+\pi^-$, non-resonant $4\pi^{\pm}$, 
$\rho^0f_0$, $f_0f_0$ and $f_0\pi^+\pi^-$ final states.
We define the likelihood function 
\begin{equation}
\mathcal{L}= \exp \biggl(-\sum_{j} n_j \biggr)
             \prod^{\mathrm{N_{cand}}}_{i=1}\biggl(\sum_{j} n_j P^i_j \biggr),
\end{equation}
where $i$ is the event identifier, 
$j$ indicates one of the event type 
categories for signals and backgrounds, 
$n_j$ denotes the yield of the $j$-th category, 
and $P^i_j$ is the probability density function (PDF)
for the $j$-th category. 
The PDFs are a product 
of two smoothed two-dimensional functions: 
$P^i_j=P_j(M^i_{\mathrm{bc}}, \Delta E^i, M^i_1, M^i_2)= 
p(M^i_{\mathrm{bc}}, \Delta E^i) \times p(M^i_1, M^i_2)$. 

For the $B$ decay components, the smoothed functions
$p_{\mathrm{smoothed}}(M^i_{\mathrm{bc}}, \Delta E^i)$ 
and $p_{\mathrm{smoothed}}(M^i_1, M^i_2)$ are 
obtained from MC simulations.
For the $M_{\mathrm{bc}}$ and $\Delta E$ PDFs, possible 
differences between real data and the MC modeling
are calibrated using a large control sample of 
$B^0 \to D^-(K^+\pi^-\pi^-)\pi^+$ decays.
The signal mode PDF is divided into two parts: 
one is correctly reconstructed events 
and the other is ``self-cross-feed'' (SCF), in which 
at least one track from the signal decay is replaced 
by one from the accompanying $B$ meson decay.
We use different PDFs for the correctly reconstructed and SCF 
events, and fix the SCF fraction to that from the MC simulation 
in the nominal fit.

For the continuum and charm $B$ decay backgrounds, 
we use the product of a linear function for $\Delta E$, 
an ARGUS function~\cite{106} for $M_{\mathrm{bc}}$
and a two-dimensional smoothed function 
for $M_1$-$M_2$.
The parameters of the linear function and ARGUS function for
the continuum events are floated in the fit.
Other parameters and the 
shape of the $M_1$-$M_2$ functions are obtained
from MC simulations and fixed in the fit.

For the charmless $B$ decay backgrounds, we use three 
separate PDFs for $B^0\to a_1^{\pm}\pi^{\mp}$,
$B^{\pm} \to\rho^{\pm} \rho^0$ and other charmless $B$ decays; 
all of the PDFs are obtained from MC simulations. 
In the fit, we fix the branching fraction of 
$B^0\to a_1^{\pm}\pi^{\mp}$ to the published value 
$(33.2\pm 3.0\pm 3.8)\times10^{-6}$~\cite{107}. 
If we float the $B^0\to a_1^{\pm}\pi^{\mp}$ yield in the fit, 
the fit result is $\mathcal{B}(B^0\to a_1^{\pm}\pi^{\mp}) = 
(33.8^{+13.4}_{-13.2})\times 10^{-6}$, which is consistent 
with the assumed value. 
We fix the yield of $B^{\pm} \to\rho^{\pm} \rho^0$ 
to that expected based on the world average branching 
fraction \cite{rho0rhop}, and we float the yield of 
other charmless $B$ decays.

\begin{figure}[htbp]
\centering
\epsfig{file=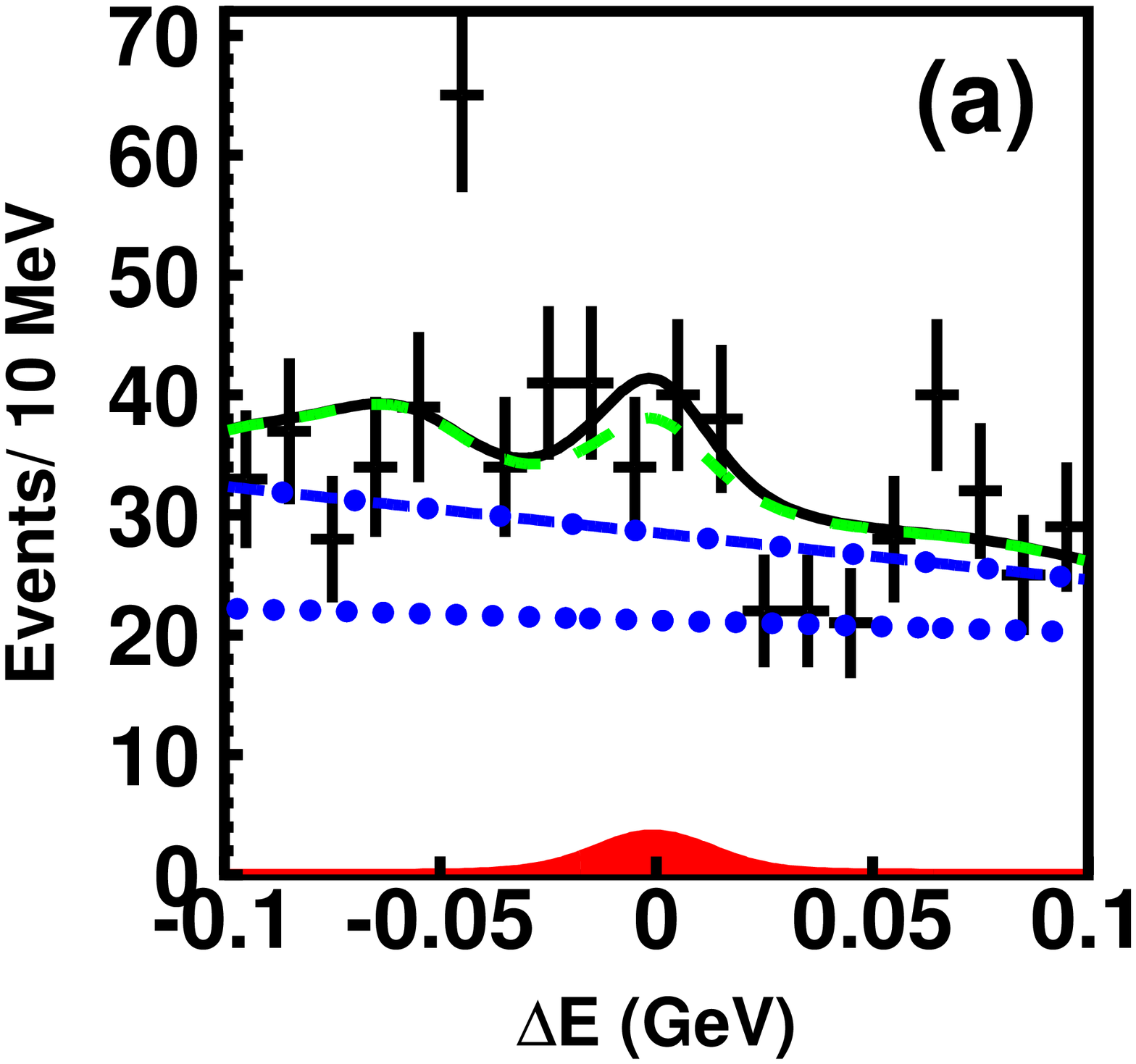,width=1.6in,height=1.6in}
\epsfig{file=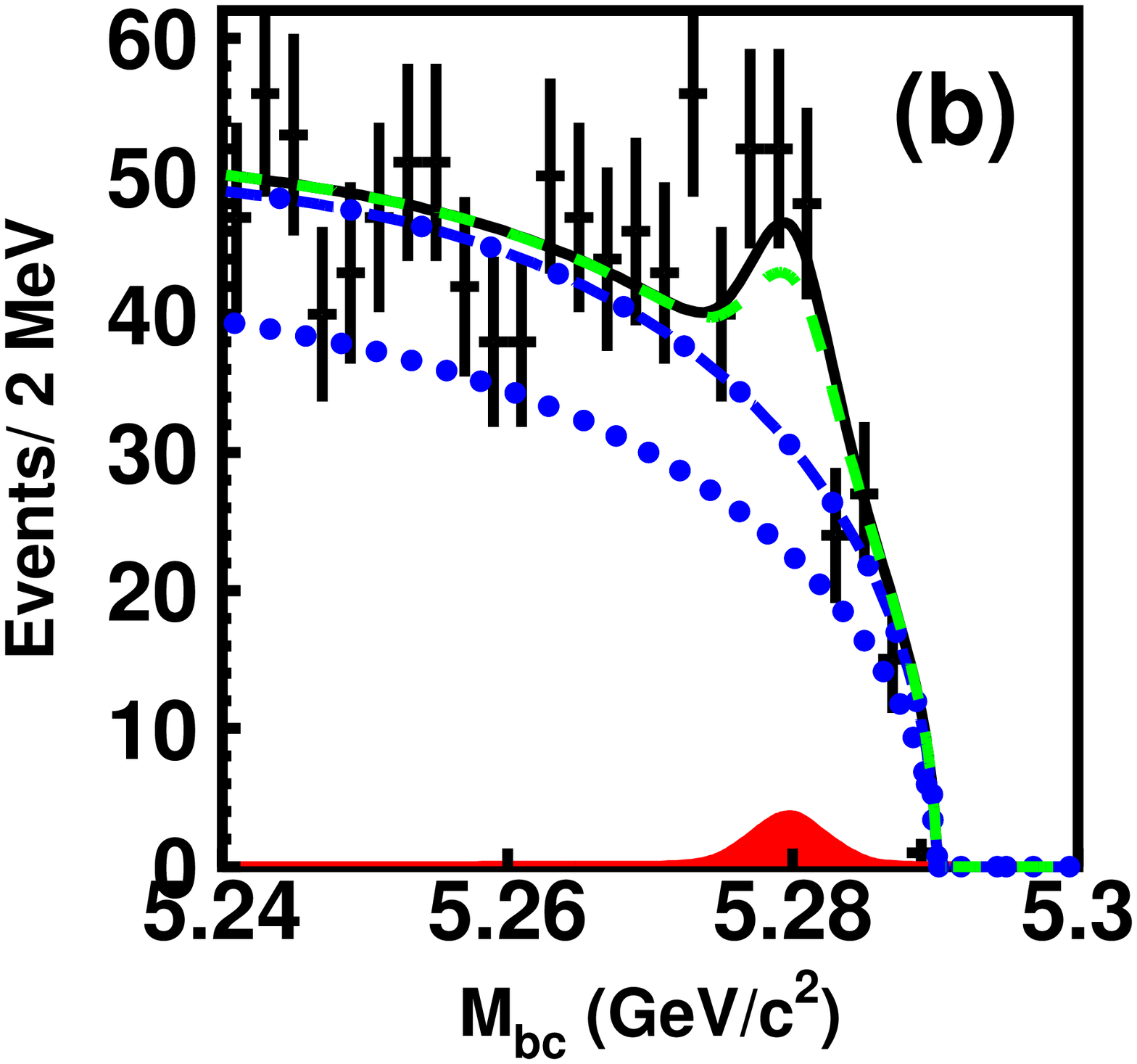,width=1.6in,height=1.6in}
\epsfig{file=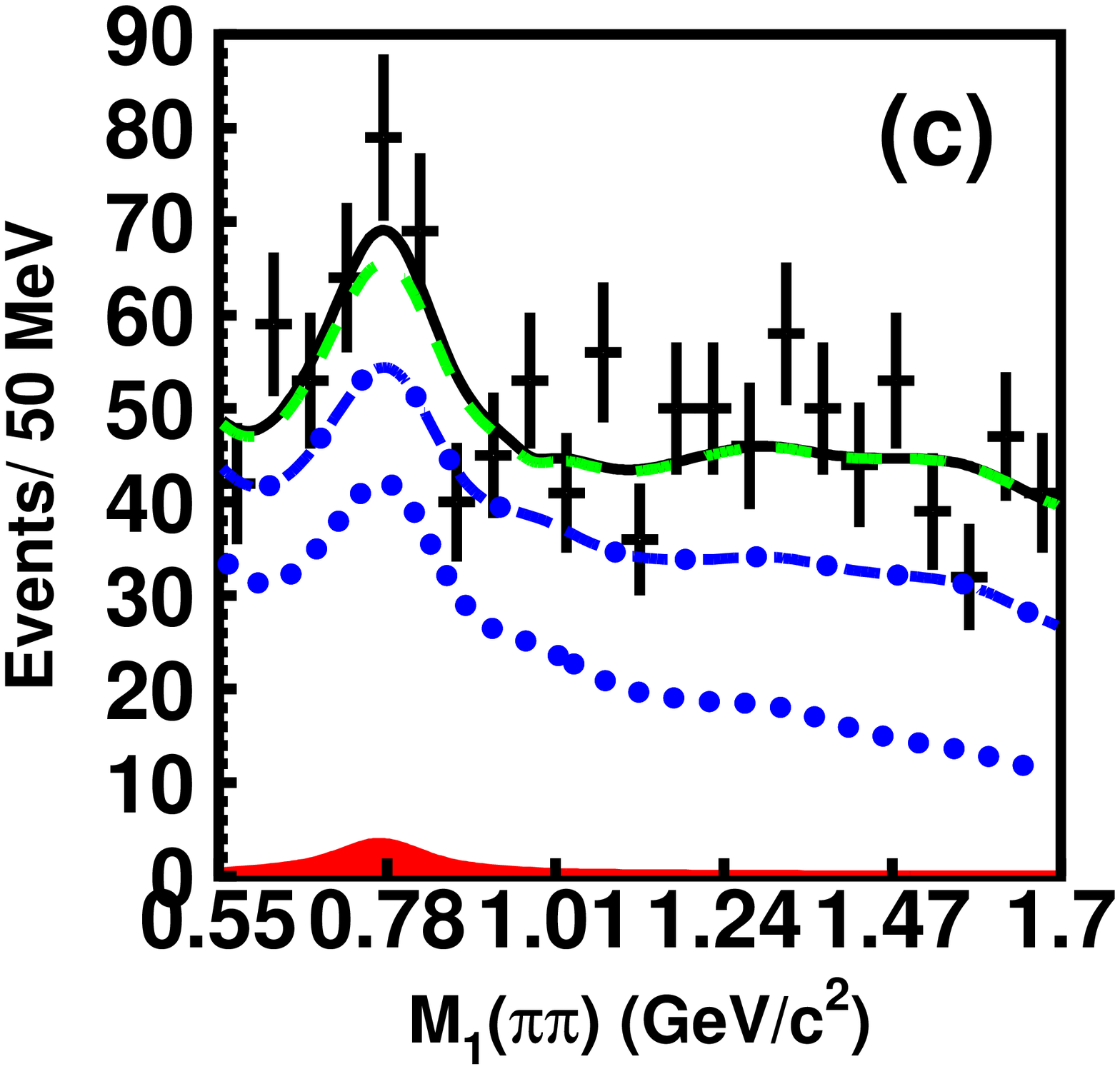,width=1.6in,height=1.6in}
\epsfig{file=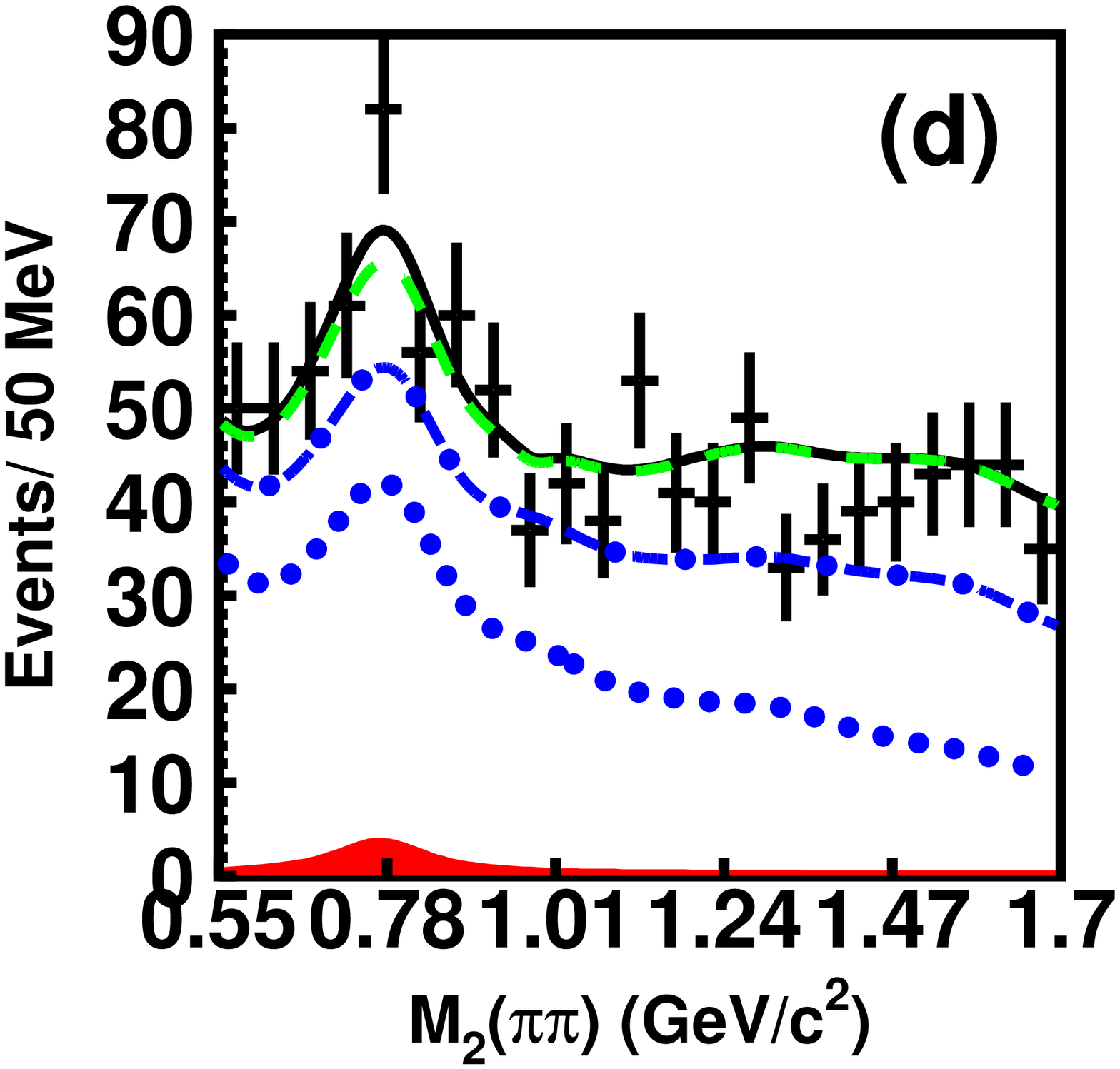,width=1.6in,height=1.6in}
\caption{
Projections of the four-dimensional fit onto (a) $\Delta E$, 
(b) $M_{\mathrm{bc}}$, (c) $M_1(\pi^+\pi^-)$, and 
(d) $M_2(\pi^+\pi^-)$, 
for candidates satisfying (except for the variable plotted) 
the criteria 
$\Delta E \in [-0.05,\ 0.05]\ \mathrm{GeV}$, 
$M_{\mathrm{bc}} \in [5.27,\ 5.29] \ \mathrm{GeV}/c^2 $, and 
$M_{1,2}(\pi^+\pi^-) \in [0.626,\ 0.926] \ \mathrm{GeV}/c^2$.
The fit result is shown as the thick solid curve; 
the solid shaded region represents the $B^0\to \rho^0\rho^0$ 
signal component. The dotted, dot-dashed and dashed curves 
represent, respectively, the cumulative background components 
from continuum processes, $b\to c$ decays, and charmless $B$ 
backgrounds.} 
\label{fig-fit1}
\end{figure}

Table~\ref{table-yield} and Fig.~\ref{fig-fit1} show the fit results and 
projections of the data onto $\Delta E$, $M_{\mathrm{bc}}$, $M_1(\pi^+\pi^-)$ 
and $M_2(\pi^+\pi^-)$ for $B^0\to\rho^0\rho^0$ decay. 
The statistical significance is defined as 
$\sqrt{-2\ln(\mathcal{L}_0 / \mathcal{L}_{\mathrm{max}})}$, 
where $\mathcal{L}_0$ and $\mathcal{L}_{\mathrm{max}}$ are the  
values of the likelihood function when the signal yield is 
fixed to zero and allowed to vary, respectively. 
The 90\% C.L. (C.L.) upper limit for the yield $N$ is 
calculated from the equation
\begin{eqnarray}
  {{\int_0^N \mathcal{L}(x) dx} \over
  {\int_0^{\infty}\mathcal{L}(x) dx} } = 90\% ,
\end{eqnarray}
where $x$ corresponds to the number of signal events. 
We include the systematic uncertainty into the upper limit (UL) 
by smearing the statistical likelihood function by a bifurcated 
Gaussian whose width is equal to the total systematic error.  
The significance including systematic uncertainties is calculated 
as before, except that we only include the additive systematic errors 
related to signal yield in the convoluted Gaussian width.

\begin{table}[htbp]
\renewcommand{\arraystretch}{1.4}
\begin{center}
\caption{Fit results for the decay modes listed in the first column. 
The signal yields, reconstruction efficiencies 
(assuming the probability for the sub-decay mode 
$f_0(980) \to \pi^+\pi^-$ is 100\%), 
significance ($\mathcal{S}$, in units of $\sigma$), 
branching fractions ($\mathcal{B}$, in units of $10^{-6}$) 
and the upper limit at the 90\% C.L. 
(UL, in units of $10^{-6}$) are listed. 
For the yields and branching fractions,
the first (second) error is statistical (systematic).}
\begin{tabular}{lccccc} 
\hline \hline
Mode               & Yield                             & Eff.(\%) & $\mathcal{S}$ & $\mathcal{B}$                 & UL \cr 
\hline 
$\rho^0\rho^0$     & $24.5^{+23.6+10.1}_{-22.1-16.2}$  & 9.16     & 1.0           & $0.4 \pm 0.4 ^{+0.2}_{-0.3}$  & $< 1.0$  \cr
$\rho^0\pi^+\pi^-$ & $112.5^{+67.4}_{-65.6} \pm 52.3$  & 2.90     & 1.3           & $5.9^{+3.5}_{-3.4} \pm 2.7$   & $< 12.0$ \cr
$4\pi^{\pm}$       & $161.2^{+61.2+27.7}_{-59.4-25.1}$ & 1.98     & 2.5           & $12.4^{+4.7+2.1}_{-4.6-1.9}$  & $< 19.3$ \cr 
$\rho^0f_0$        & $-11.8^{+14.5+4.8}_{-12.9-3.6}$   & 9.81     & $-$           & $-$                           & $< 0.3$  \cr
$f_0f_0$           & $-7.7^{+4.7}_{-3.5} \pm 3.0$      & 10.17    & $-$           & $-$                           & $< 0.1$  \cr
$f_0\pi^+\pi^-$    & $6.3^{+37.0}_{-34.7} \pm 18.0$    & 2.98     & $-$           & $0.3^{+1.9}_{-1.8} \pm 0.9$   & $< 3.8$  \cr
\hline \hline
\end{tabular}
\label{table-yield}
\end{center}
\end{table}

The fractional systematic errors are 
summarized in Table~\ref{table-sys-sum}. 
For the systematic uncertainties 
due to the fixed branching fractions, 
we vary the branching fractions of $B^0\to a_1^{\pm}\pi^{\mp}$ 
($33.2\pm 4.8$, in units of $10^{-6}$)~\cite{107} and 
$B^{\pm} \to\rho^{\pm} \rho^0$ ($18.2\pm3.0$)~\cite{108}
by their $\pm1\sigma$ errors. 
The fits are repeated and the differences between 
the results and the nominal fit values are taken as systematic errors. 
Systematic uncertainties for the $\Delta E$-$M_{\rm bc}$ PDFs used in 
the fit are estimated by performing the fits while varying the signal 
peak positions and resolutions by $\pm 1\sigma$. 
Systematic uncertainties for the $M_1$-$M_2$ PDFs 
are estimated in a similar way. 
A systematic error for the longitudinal polarization fraction of 
$B^0\to\rho^0\rho^0$ is obtained by changing the fraction from 
the nominal value $f_{\mathrm{L}}=1$ to the most extreme 
alternative value $f_{\mathrm{L}}=0$.
According to MC, the signal SCF fractions are 
20.4\% for $B^0\to\rho^0\rho^0$, 
14.2\% for $B^0\to\rho^0\pi^+\pi^-$, 
11.1\% for non-resonant $B^0\to 4\pi^{\pm}$, 
15.0\% for $B^0\to\rho^0f_0$, 
9.9\% for $B^0\to f_0f_0$ and 
13.4\% for $B^0\to f_0\pi^+\pi^-$. 
We estimate a systematic uncertainty for the signal SCF 
by varying its fraction by $\pm 50\%$. 

An MC study indicates that the fit biases are 
$+2.4$ events for $B^0\to\rho^0\rho^0$, 
$+7.2$ events for $B^0\to\rho^0\pi^+\pi^-$,
$+12.5$ events for non-resonant $B^0\to 4\pi^{\pm}$, 
$+3.6$ events for $B^0\to\rho^0f_0$, 
$-0.8$ event for $B^0\to f_0f_0$ and 
$+5.1$ events for $B^0\to f_0\pi^+\pi^-$. 
We find that fit biases occur due to the correlations between the 
two sets of variables ($\Delta E$, $M_{\rm bc}$) and ($M_1$, $M_2$), 
which are not taken into account in our fit. 
We correct the fit yields for these biases. 
To take into account possible differences between 
the MC simulation and data, we take both the 
magnitude of the bias corrections and the 
uncertainty in the corrections as systematic 
errors.

We study the possible interference between 
$B^0\to a_1^{\pm}\pi^{\mp}$, $B^0\to\rho^0\rho^0$, 
$B^0\to\rho^0\pi^+\pi^-$ and non-resonant $B^0\to 4\pi^{\pm}$ 
using toy MC. 
We add a simple interference model to the toy MC generation, 
which is, for $\rho^0\to\pi^+\pi^-$ decay, modified from 
a relativistic Breit-Wigner function to 
\begin{eqnarray}
\small
\Biggl\vert {1 \over {m^2-m_0^2+ im_0\Gamma}} 
         + A\mathrm{e}^{-i\delta}\Biggr\vert^2 = 
          \ \ \ \ \ \ \ \ \ \ \ \notag \\ 
         A^2 
           + 2A\Biggl[ {{(m^2-m_0^2)\cos\delta - \Gamma m_0\sin\delta} 
             \over {(m^2-m_0^2)^2 + (\Gamma m_0)^2 }} \Biggr] \notag \\
           + {1 \over {(m^2-m_0^2)^2+(\Gamma m_0)^2}}, 
           \ \ \ \ \ \ \ \ \ \ \ \ \ \ \ \ \ \ \ 
\end{eqnarray}
where $A$ and $\delta$ are the interfering amplitude and phase, 
and $m_0$ and $\Gamma$ are the $\rho^0$ mass and width, 
respectively. 
We assume that the interference term due to the amplitudes for 
$B^0\to a_1^{\pm}\pi^{\mp}$, $B^0\to\rho^0\pi^+\pi^-$ and 
non-resonant $B^0\to 4\pi^{\pm}$ decays are constant in 
the $B^0\to\rho^0\rho^0$ signal region. 
Since the magnitude of the interfering amplitude and relative 
phase are not known, we uniformly vary these parameters
and perform a fit in each case to measure the deviations 
from the incoherent case. 
We take the r.m.s. spread of the distribution of deviations 
as the systematic uncertainty due to interference. 

The systematic errors for the efficiency arise from the 
tracking efficiency, particle identification (PID) and  
$\mathcal{R}$ requirement. 
The systematic error on the track-finding efficiency 
is estimated to be 1.2\% per track using partially
reconstructed $D^*$ events. 
The systematic error due to the pion identification (PID) 
is 1.0\% per track as estimated using an inclusive 
$D^*$ control sample. 
The $\mathcal{R}$ requirement systematic error is 
determined from the efficiency difference between 
data and MC using a $B^0 \to D^-(K^+\pi^-\pi^-)\pi^+$ 
control sample. 
\begin{table}[htbp]
\begin{center}
\caption{Summary of systematic errors (\%) for the branching 
fraction measurements. $f_{\mathrm{L}}$ and $f_{\mathrm{SCF}}$ 
are the fractional uncertainties for longitudinal polarization 
and self-cross-feed.}
\begin{tabular}{lcccccccccc} 
\hline \hline
Source                                     & $\rho^0\rho^0$     & $\rho^0\pi^+\pi^-$ & $4\pi^{\pm}$       & $\rho^0f_0$        & $f_0f_0$           & $f_0\pi^+\pi^-$ \cr
\hline
Fitting PDF                                & $\pm$10.2          & $\pm$29.8          & $\pm$12.2          & $\pm$18.6          & $\pm$31.2          & $\pm$270        \cr
$\mathcal{B}(B^0\to a_1\pi)$               & $\pm$21.6          & $\pm$33.5          & $\pm$2.7           & $\pm$17.8          & $\pm$1.3           & $\pm$39.7       \cr
$\mathcal{B}(B^{\pm}\to \rho^0\rho^{\pm})$ & $\pm$0.0           & $\pm$0.7           & $\pm$0.2           & $\pm$0.0           & $\pm$0.0           & $\pm$1.6        \cr
$f_{\mathrm{L}}$                           & $-53.7$            & $-$                & $-$                & $-$                & $-$                & $-$             \cr
$f_{\mathrm{SCF}}$                         & $\pm$11.4          & $\pm$8.3           & $\pm$6.0           & $\pm$5.1           & $\pm$5.2           & $\pm$20.6       \cr
Fit bias                                   & $\pm$16.3          & $^{+6.4}_{-5.7}$   & $^{+7.8}_{-3.3}$   & $^{+30.5}_{-14.4}$ & $\pm$20.8          & $\pm$82.5       \cr
Interference                               & $^{+25.7}_{-20.8}$ & $-$                & $-$                & $-$                & $-$                & $-$             \cr
Tracking                                   & $\pm$5.3           & $\pm$4.6           & $\pm$4.4           & $\pm$5.0           & $\pm$4.8           & $\pm$4.5        \cr
PID                                        & $\pm$4.8           & $\pm$3.5           & $\pm$3.2           & $\pm$4.4           & $\pm$3.9           & $\pm$3.4        \cr
$\mathcal{R}$ requirement                  & $\pm$3.2           & $\pm$3.2           & $\pm$3.2           & $\pm$3.2           & $\pm$3.2           & $\pm$3.2        \cr
$N_{B \overline B}$                        & $\pm$1.4           & $\pm$1.4           & $\pm$1.4           & $\pm$1.4           & $\pm$1.4           & $\pm$1.4        \cr
\hline
Sum(\%)                                    & $^{+41.1}_{-66.0}$ & $\pm$46.5          & $^{+17.2}_{-15.6}$ & $^{+40.9}_{-30.8}$ & $\pm$38.6          & $\pm$286        \cr
\hline \hline
\end{tabular}
\label{table-sys-sum}
\end{center}
\end{table}

\begin{figure}[htbp]
\centering
\epsfig{file=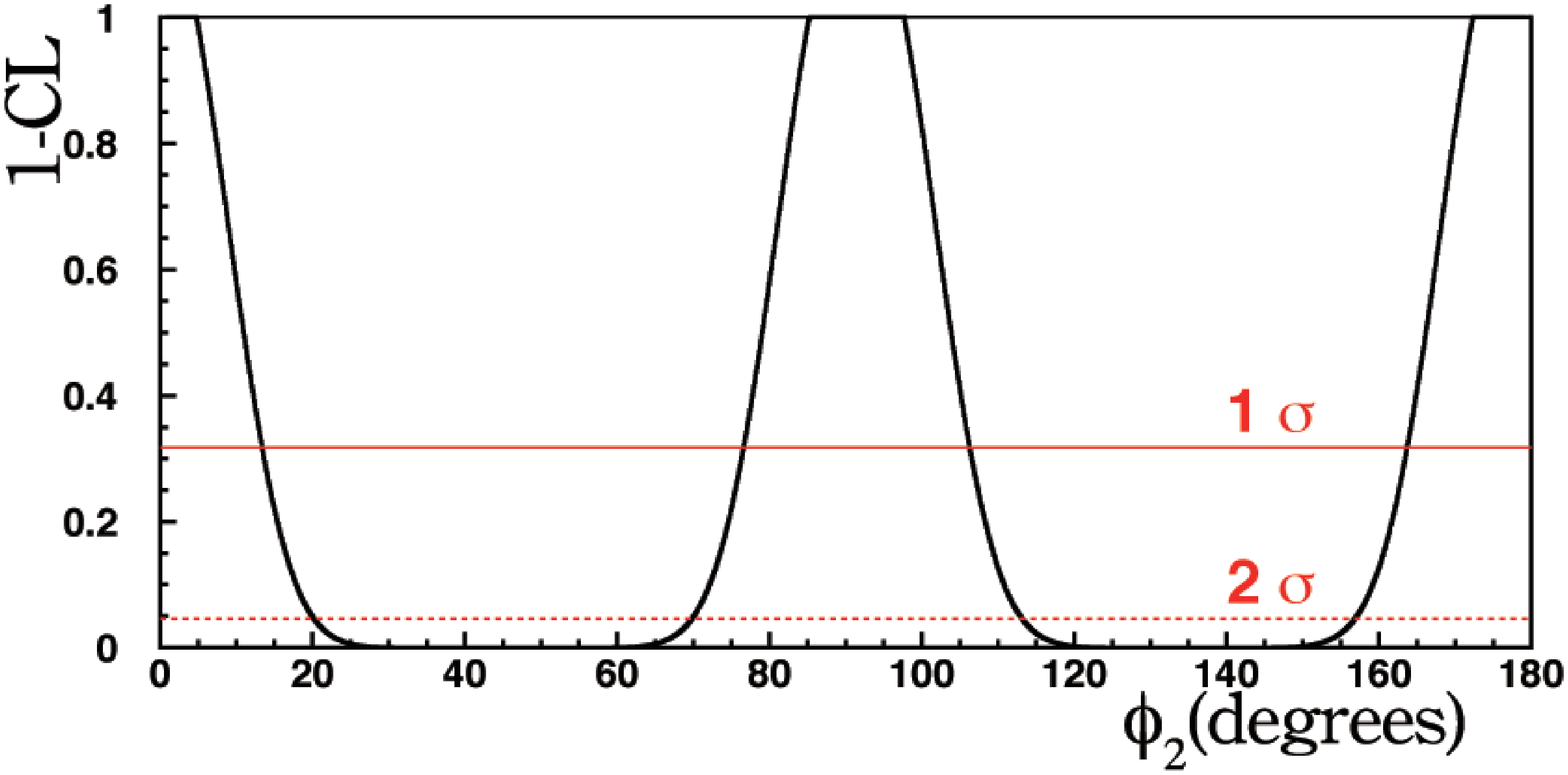,width=3.0in,height=1.8in}
\caption{1$-$C.L. vs. $\phi_2(\alpha)$ obtained from the isospin analysis 
of $B\to\rho\rho$ decays.} 
\label{fig-phi2}
\end{figure}

To constrain $\phi_2$ using $B\to\rho\rho$ decays, 
we perform an isospin analysis \cite{2, Falk} 
using the measured branching fractions of 
longitudinally polarized $B^{\pm}\to\rho^{\pm} \rho^0$, 
$B \to\rho^+ \rho^-$ and $B^0\to\rho^0\rho^0$ decays as the 
lengths of the sides of the isospin triangles.
The $B^{\pm}\to\rho^{\pm} \rho^0$ and $B \to\rho^+ \rho^-$ branching 
fractions used, as well as the corresponding $f_{\mathrm{L}}$ values, 
are world average values \cite{108}; 
the $B^0\to\rho^0\rho^0$ branching fraction is from 
this measurement, and we assume $f_{\mathrm{L}} = 1$. 
The $CP$-violating parameters $S^{+-}_L$ and 
$C^{+-}_L$ are determined from the time evolution of 
the longitudinally polarized $B \to\rho^+ \rho^-$ 
decay \cite{203, 108}.
Fig.~\ref{fig-phi2} plots the difference between one and the 
C.L. (1$-$C.L.) as a function of $\phi_2$; the 
central value and one sigma interval consistent with the SM 
is $\phi_2 = (91.7 \pm 14.9)^\circ$.

In summary, we measure the branching fraction of $B^0\to\rho^0\rho^0$ 
to be $(0.4 \pm 0.4 ^{+0.2}_{-0.3})\times 10^{-6}$ with 1.0$\sigma$
significance; the 90\% C.L. upper limit including 
systematic uncertainties is 
$\mathcal{B}(B^0\to\rho^0\rho^0)<1.0 \times 10^{-6}$. 
These values correspond to longitudinal polarization 
($f_{\mathrm{L}}=1$); the upper limit is conservative 
as the efficiency for $f_{\mathrm{L}}=1$ 
is smaller than that for $f_{\mathrm{L}}=0$. 
If we take $f_{\mathrm{L}}=0.85$,
the average of the theoretical predictions \cite{32, 13}, 
the measured value becomes $(0.3 \pm 0.3 )\times 10^{-6}$ 
(statistical error only).

On the other hand, we find excesses in $B^0\to\rho^0\pi^+\pi^-$ 
and non-resonant $B^0\to\ 4\pi^{\pm}$ decays with 1.3$\sigma$ and 
2.5$\sigma$ significance, respectively. 
We measure the branching fraction and 90\% C.L. upper limit 
for $B^0\to\rho^0\pi^+\pi^-$ decay to be 
$(5.9^{+3.5}_{-3.4} \pm 2.7)\times 10^{-6}$ and 
$\mathcal{B}(B^0\to \rho^0\pi^+\pi^-)<12.0 \times 10^{-6}$. 
For the non-resonant $B^0\to\ 4\pi^{\pm}$ mode, 
we measure its branching fraction to be 
$(12.4^{+4.7+2.1}_{-4.6-1.9})\times 10^{-6}$ with a 90\% C.L. 
upper limit of $\mathcal{B}(B^0\to 4\pi^{\pm})<19.3 \times 10^{-6}$. 
For these limits we assume the final state particles are distributed 
uniformly in three- and four-body phase space. 
We find no significant signal for the decays 
$B^0\to\rho^0f_0$, $B^0\to\ f_0f_0$ and $B^0\to\ f_0\pi^+\pi^-$; the 
final results and upper limits are listed in 
Table~\ref{table-yield}.

We thank the KEKB group for excellent operation of the
accelerator, the KEK cryogenics group for efficient solenoid
operations, and the KEK computer group and
the NII for valuable computing and SINET3 network
support.  We acknowledge support from MEXT and JSPS (Japan);
ARC and DEST (Australia); NSFC (China); 
DST (India); MOEHRD, KOSEF and KRF (Korea); 
KBN (Poland); MES and RFAAE (Russia); ARRS (Slovenia); SNSF (Switzerland); 
NSC and MOE (Taiwan); and DOE (USA).

\end{document}